%% file: Squash-arxiv.tex
\documentclass[10pt,letterpaper]{article}
\usepackage[top=0.85in,left=2.75in,footskip=0.75in]{geometry}

\usepackage{amsmath,amssymb}

\usepackage{changepage}

\usepackage[utf8x]{inputenc}

\usepackage{textcomp,marvosym}

\usepackage{cite}

\usepackage{nameref,hyperref}

\usepackage[right]{lineno}

\usepackage{microtype}
\DisableLigatures[f]{encoding = *, family = * }

\usepackage[table]{xcolor}

\usepackage{array}

\newcolumntype{+}{!{\vrule width 2pt}}

\newlength\savedwidth

\usepackage{multirow}

\usepackage{rotating}

\raggedright
\usepackage[aboveskip=1pt,labelfont=bf,labelsep=period,justification=raggedright,singlelinecheck=off]{caption}

\makeatletter
\renewcommand{\@biblabel}[1]{\quad#1.}
\makeatother

\date{}

\usepackage{lastpage,fancyhdr,graphicx}
\usepackage{epstopdf}
\fancyheadoffset[L]{2.25in}
\fancyfootoffset[L]{2.25in}
\begin{document}
\vspace*{0.2in}

\begin{flushleft}
{\Large
\textbf\newline{Audio-based performance evaluation of squash players} 
}
\newline
\\
Katalin Hajd\'u-Sz\"ucs\textsuperscript{1*},
N\'ora Fenyvesi\textsuperscript{2},
J\'ozsef St\'eger\textsuperscript{2},
G\'abor Vattay\textsuperscript{2}
\\
\bigskip
\textbf{1} Dept. of Information Systems, E\"otv\"os Lor\'and University, Budapest, Hungary
\\
\textbf{2} Dept. of Physics of Complex Systems, E\"otv\"os Lor\'and University, Budapest, Hungary
\\
\bigskip

* szucsk@caesar.elte.hu

\end{flushleft}
\section*{Abstract}
\input{abstract}


\section{Introduction}
\label{Introduction}

At present in competitive sports there are a lot of talented sportsmen and the differences between individual performance are often very small to spot.
It catalyses a race condition to be present already in the practising period, thus more and more coaches and players seek finding different means and aids to elaborate and make the preparation for the tournaments always more effective.
There are a lot of new technological achievements available in the market.
Small electronic devices are capable of measuring various metrics including those that are relevant for the sports, like heart rate and blood temperature and pressure registers, pedometers, speedometers  and accelerometers to name a few.
Using such devices is more than necessary since the results in a competition and then the final scores may depend on millesimal of millimeters. 
Another reason why to use measurement devices yielding objective performance metrics is because when sportsmen are overloaded in a performance, with adrenalin in their vein, it is hard if possible for them to spot and fix their failures.
In certain types of sports a continuous or prompt feedback is definitely helpful, squash is one of them.

Squash is a very rapid ball and racquet game with typically 40-60 hit events per second.
Depending on the various surfaces the ball interacts during its flight defines the different shot classes.
Some shot classes are very rare due to being tricky to deliver or may occur only in circumstances where the rally may seem already lost.
So knowing the detailed statistics of various hits and shot patterns talks about the quality of the sportsmen and are very important information for both the coaches and the squash players. 
However, these data and their statistical analysis are not available at present because of the paste of squash. 
Given its fast speed the human processing of events enables the score registration in real-time only, but the recording of shot types and the detailed sequences of the shots are rendered definitely impossible. 
One possible solution might be to analyse videos of the matches using image processing as it has been shown to work for the tennis~\cite{broadbent_tenniscam}.
Though for the squash it turns out that this approach remains difficult even with the use of high speed and high resolution cameras, due to the small size of the ball and the view provided by the cameras.
Traditionally cameras are placed behind the court, therefore the players will most often cover the sight of the ball during the match making the reconstruction of ball trajectories an inauspicious problem.
To provide reliable statistics by this approach will require human processing and validation so in the end a thorough analysis of the tournament will cost many times of the duration of this sport events in man-hours.

In this study we introduce a framework to unhide these information based on the analysis of acoustic data.
Playing squash produces characteristic sound patterns.
The sound footprint of each rally is a projection of all the details about the strength and the position of the ball hitting various surfaces in the court.
Naturally, this pattern, which maintains the natural order of the events, is contaminated by some additional noise.
Recording the sound in more directions allows for inverting the problem and for giving statistical statements about where and what type of an events took place in the play.
We are focusing on events generated by the ball hits, which serves as a basis for further analysis and the reconstruction of shot patterns or the ball trajectories.
Note, the framework to be detailed can be applied to various other types of ball games.

The subsequent sections of the paper are structured as follows. 
Section~\ref{sec:equipment} details the hardware components installed in a squash court to record input.
In sections \ref{sec:BID}, \ref{sec:localization} and \ref{sec:classification} mathematical models are presented to detect, localize and classify audio events respectively.
The data collection is described and the results are presented in section~\ref{sec:results}. 
Finally, methods described in this study is compared to the related works of the topic in section~\ref{sec:relwork}.

\section{The measurement equipment}
\label{sec:equipment}

This study is based on the analysis of sound waves generated during the squash play.
Among many other, squash is a game where various different sources of sound are present, including the players themselves (their sighing or their shoes squeaking on the floor), the ball hitting surfaces (like the walls, the floor or the racquet) and also external sources (including the ovation of the spectators or sound generated in an adjacent court).
Here we focus on audio events related to the ball.

When planning the experiments the following constraints had to be investigated and satisfied.
The framework should be fast in signal processing point of view, because the target information can be most valuable when in a competitive situation it helps fine tune tactical decisions made by the coach and/or the player.
The cost of the equipment should be kept low and the installation of the sensors requires a careful design to prevent them from disrupting the play.
As the spatial localization of the ball is one of the fundamental goals a lower bound to the sampling rate is enforced to remain able to differentiate between displaced sound sources.

\begin{figure}[!h]
\includegraphics[width=0.9\textwidth]{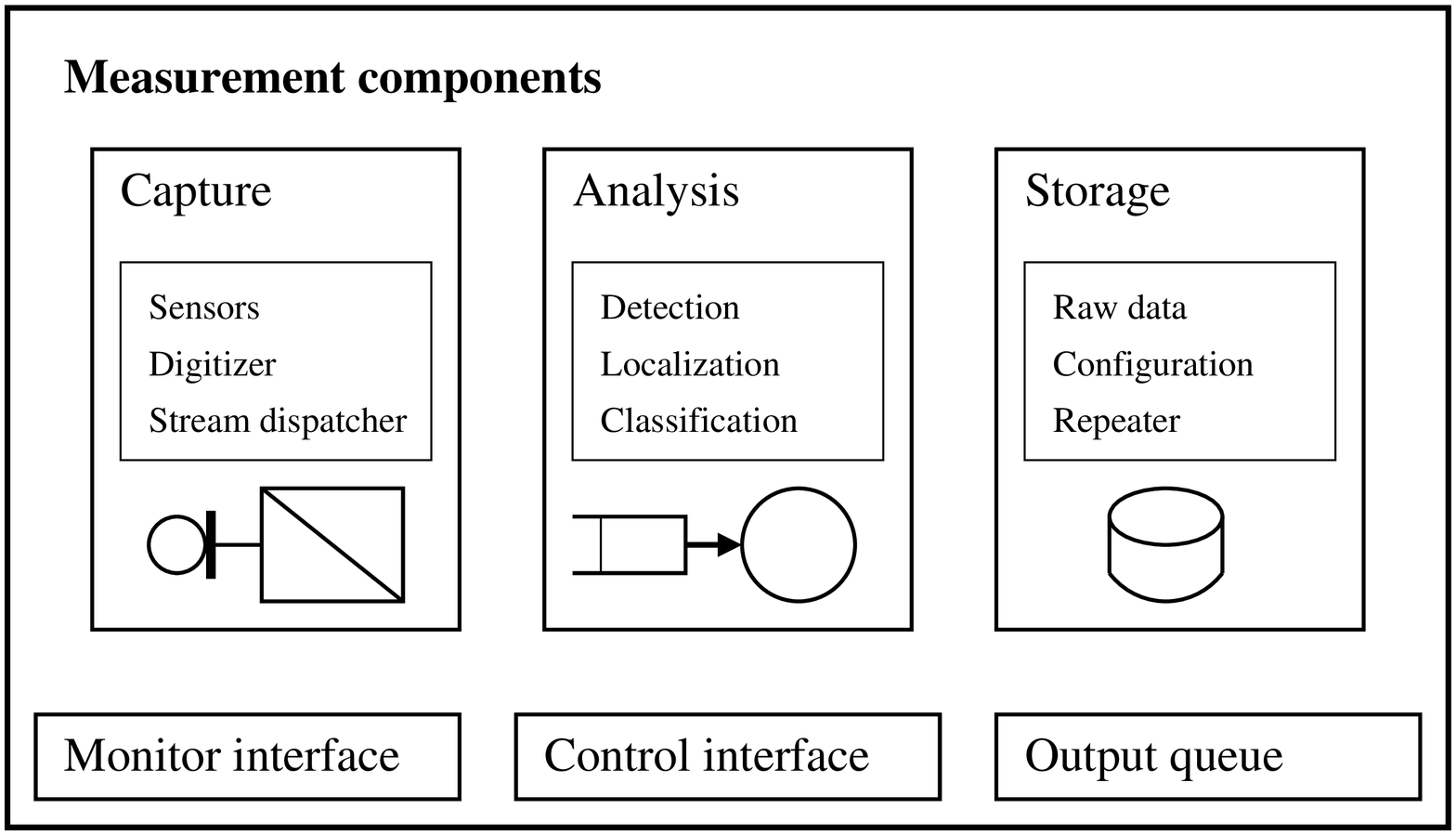}
\caption{{\bf A schematic view of the components.}
To process audio events in the sqash court a three component architecture was designed. 
}
\label{fig:overall}
\end{figure}

In Fig~\ref{fig:overall} the hardware and software components are sketched.
Hardware components include 6~audio sensors, three of which are omnidirectional microphones (Audio Technica ES945) sinking in the floor and the rest of them are cardioid microphones (Audio Technica PRO 45) hanging from the top.
Amplification and sampling of the microphone signals are done by a single dedicated sound card (Presonus AudioBox 1818VSL) so that all channels in a sample frame are in synchrony.
The highest sampling rate of the sound card is used (96~kHz), so by each new sample the front of a sound wave travels approximately 6~mm.

According to their functionalities software components fall in the following groups.
Signal processing is done in the analysis module, which include the detection of the audio events, the classification and the filtering of the detections and after matching event detections of more channels the localization of the sound source.
While these steps of signal processing can be done real-time a storage module is also implemented so that the audio of important matches can be recorded.
Recording of data helps training of the parameters of the classification algorithms, and it also enables a whole re-analysis of former data with different detectors and/or different classifiers.
All output generated by the Analysis module is fed to the output queue.
Hardware and software components are triggered and reconfigured via a web services API exposed by the Control interface.
Finally, to be able to listen to what is going on in the remote court a Monitoring interface provides a mixed, downsampled and compressed live stream across the web.

\section{The ball impact detection}
\label{sec:BID}
The localization and the classification of ball hits both require the precise identification of the beginning of the corresponding events in the audio streams.
The detection of ball impact events is carried out for each audio channels independently and in a parallel fashion, which speeds up the overall performance of the framework significantly.
Different detection algorithms of various complexities were investigated two extreme cases are sketched here.
The first model assumes that the background noise follows the normal distribution. 
An event is detected if new input samples deviate from the Gaussian distribution to a certain predefined threshold value.
Next for each channels the mean and the variance estimates of a finite subset of the samples are continually updated according to the Welford's algorithm~\cite{welfordvariance}.

The second method is an extension of the windowed Gaussian surprise detection by Schauerte and Stiefelhagen~\cite{BayesianSurprise}. 
The algorithm tackles the problem evaluating the relative entropy~\cite{kullback}.
It is first applied in the frequency domain and if there is a detection then a finer scale search is carried out in the time domain.
The power spectrum of $w$-sized chunks of windowed data samples is calculated.
Between detection regime the series of the power spectra is modelled by a $w$-dimensional Gaussian.
The a priori parameters of the distribution are calculated for $n$ elements in the past, and the posteriori parameters are approximated including the new power spectrum.
The Kullback Leibler divergence between the a priori and the posteriori distributions exceeds a predefined threshold when a new detection takes place
{\small\[
\mathrm{S}_i = \frac{1}{2}\left[
\log \frac{|\Sigma_i|}{|\Sigma'_i|}
+ \text{Tr}\left(\Sigma_i^{-1} \Sigma'_i\right) - w + 
\left(\mu'_i - \mu_i\right)^T \Sigma_i^{-1} \left(\mu'_i - \mu_i\right)
\right],
\]}%
where primed parameters correspond to the posteriori distribution.
The time resolution at this stage is $w$ and to increase precision a new search is carried out in the time domain evaluating the Kullback Leibler divergence for 1-d data.
In order to bootstrap a priori distribution parameters $n$ samples from the former windows are used.

%
\section{The localization of sound events}
\label{sec:localization}
In this section we lay down a probabilistic model to determine the time and location of an audio event.
For a unique event we denote these unknowns $t$ and $\mathbf{r}_\text{ev}$ respectively.
The inputs required to find the audio event are the locations of the $N+1$ detectors $\mathbf{r}_i^\text{mike}$ and the timestamps $\tau_i$ when these synchronized detectors sense the event ($0\leq i\leq N$).

The probability that microphone $i$ detects an event at $(\mathbf{r}, t)$ is
$$p(t_i, r_i) = \frac{1}{\sqrt{2\pi}\sigma_i} \exp -\frac{(ct_i-r_i)^2}{2\sigma_i^2 c^2},$$
where $c$ is the speed of sound, $t_i = \tau_i - t$ is the propagation delay and $r_i = ||\mathbf{r}-\mathbf{r}_i^\text{mike}||$ is the distance between the sound source and the microphone.
The uncertainty $\sigma_i$ depends on the characteristics of the microphone, which we will consider constant in the first approximation.

By introducing relative delays $\hat\tau_i=\tau_i-\tau_0$ the joint probability of relative delays detected is
$$p(\hat\tau_1,\dots\hat\tau_N)=\int\mathrm{d}t_0\, p(t_0, r_0) \prod_{i=1}^{N} p(\hat\tau_i+t_0, r_i).$$
The formula can be rearranged
$$p(\hat\tau_1,\dots\hat\tau_N)=\frac{1}{\sqrt{2\pi}^{N+1}\prod_{i=0}^N\sigma_i}\int\mathrm{d}t_0\, e^{-f(t_0)},$$
where $f(t_0)=\sum_{i=0}^N \frac{(c\hat\tau_i + c t_0 - r_i)^2}{2\sigma_i^2 c^2}$ is a quadratic function and in the expression for $p$ the Gaussian integral follows
$$\int\mathrm{d}t_0\, e^{-f(t_0)}=\sqrt{\frac{2\pi}{f^{\prime\prime}(t^*_0)}}e^{-f(t^*_0)}.$$
The first order derivative $f^{\prime}$ vanishes in $t^*_0=\Sigma^2\,\sum_{i=0}^N\frac{1}{\sigma_i^2}\left(\frac{r_i}{c}-\hat\tau_i\right)$, where $\Sigma^2=1/\sum_{i=0}^N\frac{1}{\sigma_i^2}$ is introduced for convenience.

After substitution of $t^*_0$ we arrive at
$$f(t^*_0)=\frac{1}{2}\left\{\sum_{i=0}^N\frac{1}{\sigma_i^2}\left(\frac{r_i}{c}-\hat\tau_i\right)^2 - \Sigma^2 \left[\sum_{i=0}^N\frac{1}{\sigma_i^2}\left(\frac{r_i}{c}-\hat\tau_i\right) \right]^2 \right\}.$$
This formula can be interpreted as a variance formula, which can be rewritten
$$f(t^*_0)=\frac{1}{2\Sigma^2} \sum_{i=0}^N\frac{1}{\sigma_i^2} \left[\sum_{j=0}^N\frac{1}{\sigma_j^2} \left(\frac{r_i-r_j}{c}-(\hat\tau_i-\hat\tau_j)\right)\right]^2.$$

A good approximation of the audio event maximizes the likelihood $p$, which at the same time minimizes $f(t^*_0)$, thus we seek the solution of $\nabla_\mathbf{r}f(t^*_0)=0$ equations.

\begin{figure}[!h]
\includegraphics[width=0.9\textwidth]{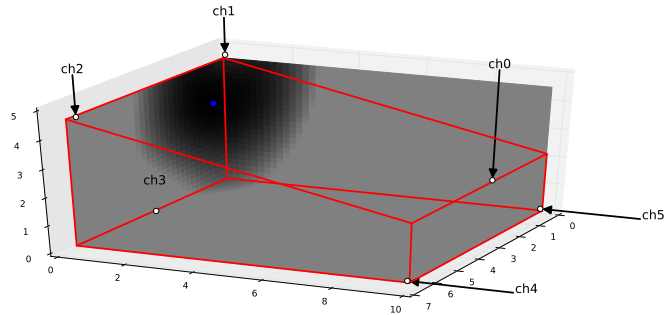}
\caption{{\bf The visualization of the likelihood function.}
The ball hit the front wall. $f(t^*_0)$ can be evaluated in space given the positions of the sensors (marked by white disks) to find its minimum, which indicates where the event took place. (0.5~m from the right corner and 3~m above the floor, marked by a blue disk.)}
\label{fig:likelihood}
\end{figure}

In practice $f$ behaves well and its minimum can be found by gradient descent method.
Fig~\ref{fig:likelihood} shows a situation, where the ball hit the front wall and 6 microphones detect this event error free.
To show the functions behaviour $f$ is evaluated in the floor, in the front wall and in the right side wall.
Finding the minimum of $f$ takes less than ten gradient steps.

\begin{figure}[!h]
\includegraphics[width=0.9\textwidth]{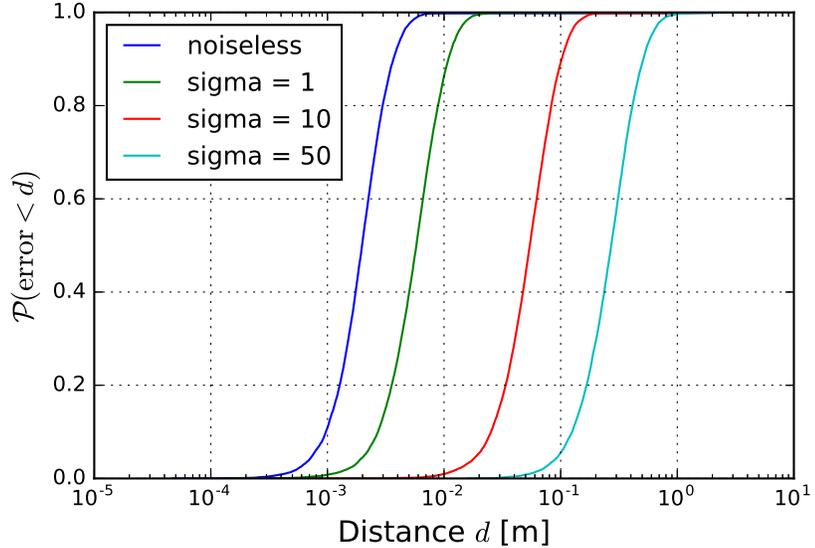} 
\caption[error]{{\bf The cumulative distribution of the localization error.}
For a noiseless case most often localization will have an error comparable to the size of the ball. With a bad detector ($\sigma=50$ samples) still the localization is exact in the order of 10~cm.}
\label{fig:errorpropagation}
\end{figure}

The likelihood based localization model is derived for a noiseless situation, assuming the perfect detection of samples in each channel.
In real environment, however, noise is present and the error deviating the detection is exposed in the final result of the localization.
In order to track this effect the method was numerically investigated as follows.
10000 points in the volume of the court is selected randomly and the sound propagation is calculated in each six microphones.
Next for the ideal detections Gaussian noise is added in all channels, with increasing variation ($\sigma=1, 10, 50$).
In Fig~\ref{fig:errorpropagation} the noiseless case is compared to cases with increasing errors.
In the figure the cumulative distribution of the error, ie. the difference between the randomly selected point and the location guess by the model is presented.
Naturally, by increasing the detection error the error in the position guess is increasing, but the model performs very well, for poor signal detectors the error in localization is in the order of 10~cm.

\section{Classification}
\label{sec:classification}

It is the task of the classification module to distinguish between the different sound events according to their origin.
Sound events are classified based on the type of the surface that suffered from the impact of the ball. 
This surface can be the wall, the racquet, the floor or the glass. 
When the sound does not fit any of these classes, like the squeaking shoes, then it is classified as a false event. 
The classification enhances the overall performance of the system by two means. 
First, skipping to localize the false events speeds up the processing.
And second, in doubtful situations when the calculated location of the event falls near to multiple possible surfaces, by knowing the type of the surface that suffered from the impact can reinforce the localization.
For example a sound event localized a few centimetres above the floor could be generated by a racquet hit close to the floor or by the floor itself.

Classification utilizes feed-forward neural networks that had been trained with backpropagation~\cite{Hinton:2012, bugatti2002audio,shao2003applying,wang2007sound}.
The training sets are composed of vectors belonging to 5461 audio events, which have been manually labelled.
Based on these audio events two types of input were constructed for teaching.

In the first case temporal data is used directly.
A vector element of the training set $T_1$ is the sequence of the samples around the detections for each channels.
\[
T_1 = \left\{ (a_{d-w},\dots,a_d,\dots,a_{d+w}) \right\},
\]
where the channel index is dropped and $d$ is a uniq detection and $w$ sets the length of the vector.
Given the sampling rate 96~kHz  and setting $w=300$ the neural network is taught by $6.25$~millisecond long data.

The second feature set $T_2$ is built up of the power spectra.
\[
T_2 = \left\{ |\mathcal{F}(a_{d},\dots,a_{d+w})| \right\},
\]
where $\mathcal{F}$ denotes the discrete Fourier transform.

A single neural network model where all event classes are handled together performed poorly in our case.
Therefore, separate discriminative neural network models were built for all four classes (racquet, wall, floor and glass impact) and for both of the training sets.
It has also been investigated if any of the input channels introduce discrepancy.
In order to discover this effect models were built and trained for each unique channels and another one handling the six channels together.
Note, that not all possible combinations of the models were trained due to the fact that some channels poorly detected certain events, for example microphones near the front wall detected glass events very rarely.

In the training sets the class of interest was always under-represented.
To balance the classifier the SMOTE~\cite{chawla2002smote} algorithm was used, which is a synthetic minority over-sampling technique. 
A new element is synthesized as follows.
The difference between a feature vector from the positive class and one of its $k$ nearest neighbours is computed.
The difference is blown by a random number between 0 and 1, to be added to the original feature vector. 
This technique forces the minority class to become more general, and as a result, the class of interest becomes equally represented like the majority set in the training data.

Different network configurations were realized to find that for the direct temporal input a 20 hidden layer network (with 10 neurons in each layer) performed the best, while for the spectra input a 10 hidden layer (each layer with 10 neurons) is the best choice.

\section{Analysis}
\label{sec:results}
In this section the performance of each modules of the framework and the datasets are presented.

\subsection{Datasets}
\label{sec:datasets}
In order to analyse the components of the framework implementing the proposed methods 
two audio record sets were used.
\emph{Audio 1} was recorded on the 18th of May 2016 when a squash player was asked to target front wall shots to specific areas of the wall. 
This measurement was necessary to increase the cardinality of front wall and racquet hit significantly in the training datasets $T_1$ and $T_2$, and it was also manually processed to be able to validate the operations of the detector and the localization components.
\emph{Audio 2} resembles data in a real situation as it contains a seven minutes squash match recorded on the 8th of March 2016. 
Table~\ref{table:audio} summarizes the details of these audio recordings.

\begin{table}[!ht]
\centering
\caption{{\bf The content of the audio files.}}
\begin{tabular}{|c|l|r|r|r|r|r|r|r|}
\hline
& Class & Ch0& Ch1& Ch2& Ch3& Ch4& Ch5& \cellcolor{gray!20!white} Total \\
\hline
\multirow{3}{*}{\begin{turn}{90}Audio 1\end{turn}} & Front wall& 165& 165& 165& 165& 165& 165& \cellcolor{gray!15!white} 990 \\
& Racquet& 166& 166& 166& 166& 166& 166& \cellcolor{gray!15!white} 996 \\
& \cellcolor{gray!15!white} Total& \cellcolor{gray!15!white} 331& \cellcolor{gray!15!white} 331& \cellcolor{gray!15!white} 331& \cellcolor{gray!15!white} 331& \cellcolor{gray!15!white} 331& \cellcolor{gray!15!white} 331& \cellcolor{gray!15!white}1986 \\
\hline
\multirow{7}{*}{\begin{turn}{90}Audio 2\end{turn}} & Front wall& 100& 109& 108& 110& 107& 111& \cellcolor{gray!15!white} 645 \\
& Racquet& 112& 112& 113& 110& 109& 99& \cellcolor{gray!15!white} 655 \\
& Floor& 85& 70& 75& 19& 115& 11& \cellcolor{gray!15!white} 375 \\
& Glass& 46& 20& 24& 15& 62& 11&  \cellcolor{gray!15!white} 178 \\
& False event & 227& 274& 254& 264& 456& 147& \cellcolor{gray!15!white} 1622 \\
& \cellcolor{gray!20!white} Total& \cellcolor{gray!15!white} 570& \cellcolor{gray!15!white} 585&  \cellcolor{gray!15!white} 574 & \cellcolor{gray!15!white} 518& \cellcolor{gray!15!white} 849& \cellcolor{gray!15!white} 379&  \cellcolor{gray!15!white} 3475\\
\hline
\end{tabular}
\begin{flushleft}
The count of events in \emph{Audio 1} and \emph{Audio 2} broken down for each class and each channel. In total 5461 events have been labeled.
\end{flushleft}
\label{table:audio}
\end{table}

Training the neural network models require properly labelled datasets.
After applying the ball impact detection algorithm to the audio records the timestamps of the detected events were manually categorized as front wall event, racquet event, floor event, glass event or false event.

\subsection{Detection Results}
\label{sec:resdetector}

\begin{figure}[!h]
\includegraphics[width=0.9\textwidth]{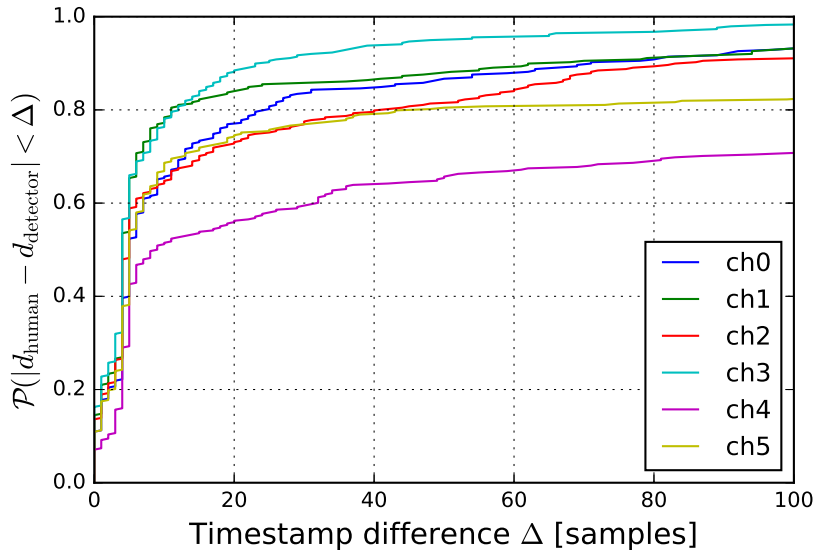}
\caption{{\bf The error of the detector.}
The detection error is defined as the difference between the timestamps generated by the module and read by a human.}
\label{fig:deterr}
\end{figure}

The performance of the detector is analysed by comparing the timestamp reported by the detector $d_\mathrm{detector}$ and the human readings $d_\mathrm{human}$.
For \emph{Audio 1} in Fig~\ref{fig:deterr} the cumulative probability distribution of the time difference is shown for each channel and in Table~\ref{table:deterrch} the average error and its variance are shown grouped by the two event types present in the dataset. 
One can observe that the detectors in channels \emph{ch4} and \emph{ch5} perform poorly.
When estimating the position discarding one of or both of these channels will enhance the precision of the localization.

\begin{table}[!ht]
\centering
\caption{{\bf The class and channelwise error of the detector.}}
\begin{tabular}{|l|c|c|}
\hline
 & Front wall & Racquet \\
\hline
ch0 &    9.6 $\pm$  46.0 &  -5.8 $\pm$  63.7 \\
ch1 &    3.1 $\pm$   1.9 &  -9.3 $\pm$ 130.6 \\
ch2 &    3.5 $\pm$   5.4 &  21.3 $\pm$ 129.3 \\
ch3 &    3.0 $\pm$   1.9 &   7.3 $\pm$  39.9 \\
ch4 &  221.4 $\pm$ 476.5 & 116.4 $\pm$ 401.3 \\
ch5 &  210.8 $\pm$ 512.3 &  23.5 $\pm$ 136.2 \\
\hline
\end{tabular}
\begin{flushleft}
The error of the detector algorithm is measured in samples for the various classes and all channels.
\end{flushleft}
\label{table:deterrch}
\end{table}

In Table~\ref{table:deterrtype} the error statistics for dataset \emph{Audio 2} is shown. 
Intensive events, like front wall impacts, can be detected precisely, whereas the detection of milder sounds like a floor or glass impact is less accurate.
\begin{table}[!ht]
\centering
\caption{{\bf Classwise error of the detector.}}
\begin{tabular}{|l|c|c|c|}
\hline
 Class     & Audio 1 & Audio 2 \\
\hline
Front wall &  4.8 $\pm$  23.3 &   6.9 $\pm$  19 \\
Racquet    &  3.4 $\pm$  99.8 & 107   $\pm$  85 \\
Floor      & 38.0 $\pm$ 141.1 & 125   $\pm$ 149 \\
Glass      & \emph{n.a.}  & 183   $\pm$ 173 \\
\hline
\end{tabular}
\begin{flushleft}
The statistics of the dataset \emph{Audio 1} is calculated for 660 events for each class excluding Floor events, counting 24 pieces. For \emph{Audio 2} 200 events were available for each class.
\end{flushleft}
\label{table:deterrtype}
\end{table}

The false discovery and the false negative rate of the detector were examined on \emph{Audio 2}.
False positives are counted if detector signals for a false event, and false negatives are the missing detections.
The results are summarised in Table~\ref{table:detectorconfusion}.
\begin{table}[!ht]
\centering
\caption{{\bf Performance of the detector.}}
\begin{tabular}{|c|r|r|r|r|r|r|}
\hline
False alarm & Ch0 & Ch1 & Ch2 & Ch3& Ch4 & Ch5 \\
\hline
FDR & 39\% & 47\% & 44\% & 51\% & 54\%& 39\% \\
FNR& 16\%& 24\%& 22\%& 38\%& 5\% & 43\% \\ 
\hline
\end{tabular}
\begin{flushleft}
False Discovery Rate (FDR: $\frac{n_\mathrm{fp}}{n_\mathrm{tp} + n_\mathrm{fp}}$) and False Negative Rate (FNR:$\frac{n_\mathrm{fn}}{n_\mathrm{fn} + n_\mathrm{tp}}$) of the detector based on 3475 events.
\end{flushleft}
\label{table:detectorconfusion}
\end{table}

\subsection{Classification Results}
\label{CP}

Approaching the problem at first and to use as much information as possible to teach the neural networks a large training set was constructed of the union of the detections of all the six channels.
However, this technique gave poorer results than treating all the channels separately. The different settings of the microphones and the distinct acoustic properties of the squash court at the microphone positions are found to be the reasons of that phenomenon.

\begin{figure}[!h]
\centering
\includegraphics[width=.9\textwidth]{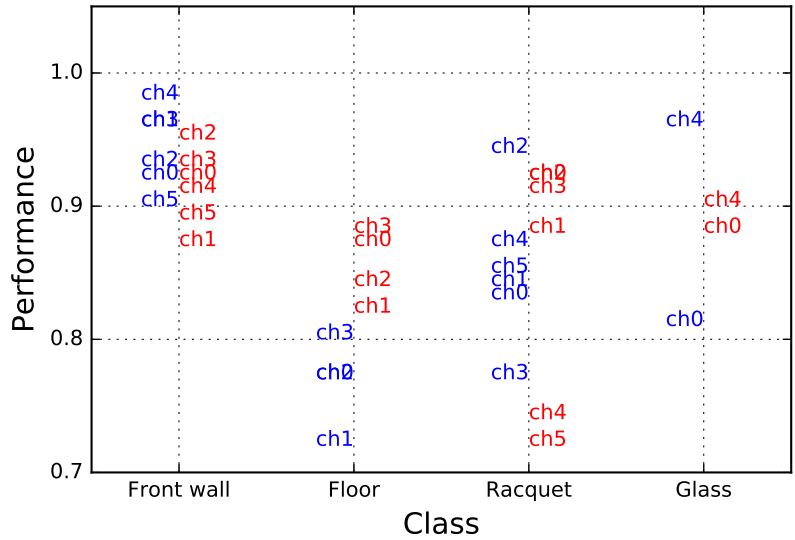}
\caption{{\bf The classifiers' accuracy.}
The classwise accuracy of each channel is presented in $T_1$ (blue) and $T_2$ (red) input sets. 
Front wall classification gives high accuracy on all channels in both sets. 
It is interesting to observe that floor classification is more accurate in input $T_2$. 
Racquet classification performs best on channel 2 in both sets.}
\label{fig:cl_acc}
\end{figure}

\begin{figure}[!h]
\centering
\includegraphics[width=.9\textwidth]{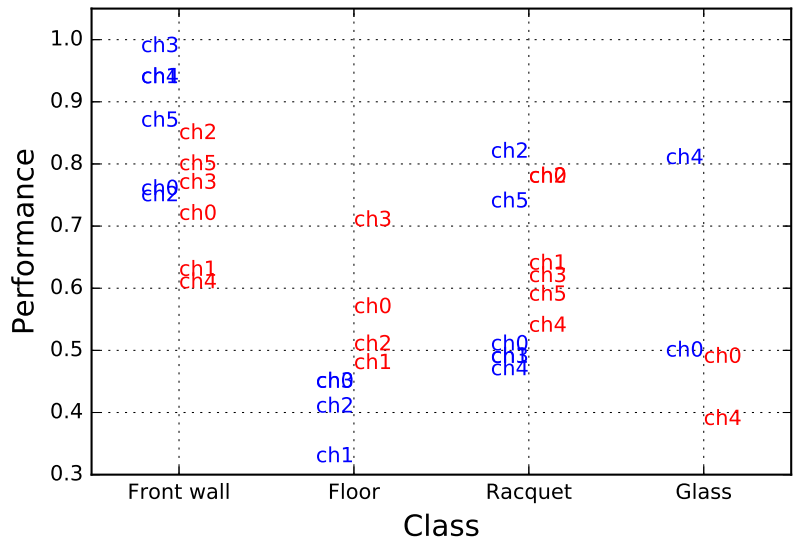}
\caption{{\bf The classifiers' precision.} 
The classwise precision of each channel is presented in $T_1$ (blue) and $T_2$ (red) input sets. 
Front wall classification gives high precision in input $T_1$. 
The precision of floor classification is low. 
Racquet classification still performs best on channel 2. 
The precision of glass classification is only acceptable on channel 4.}
\label{fig:cl_prec}
\end{figure}

\begin{figure}[!h]
\centering
\includegraphics[width=.9\textwidth]{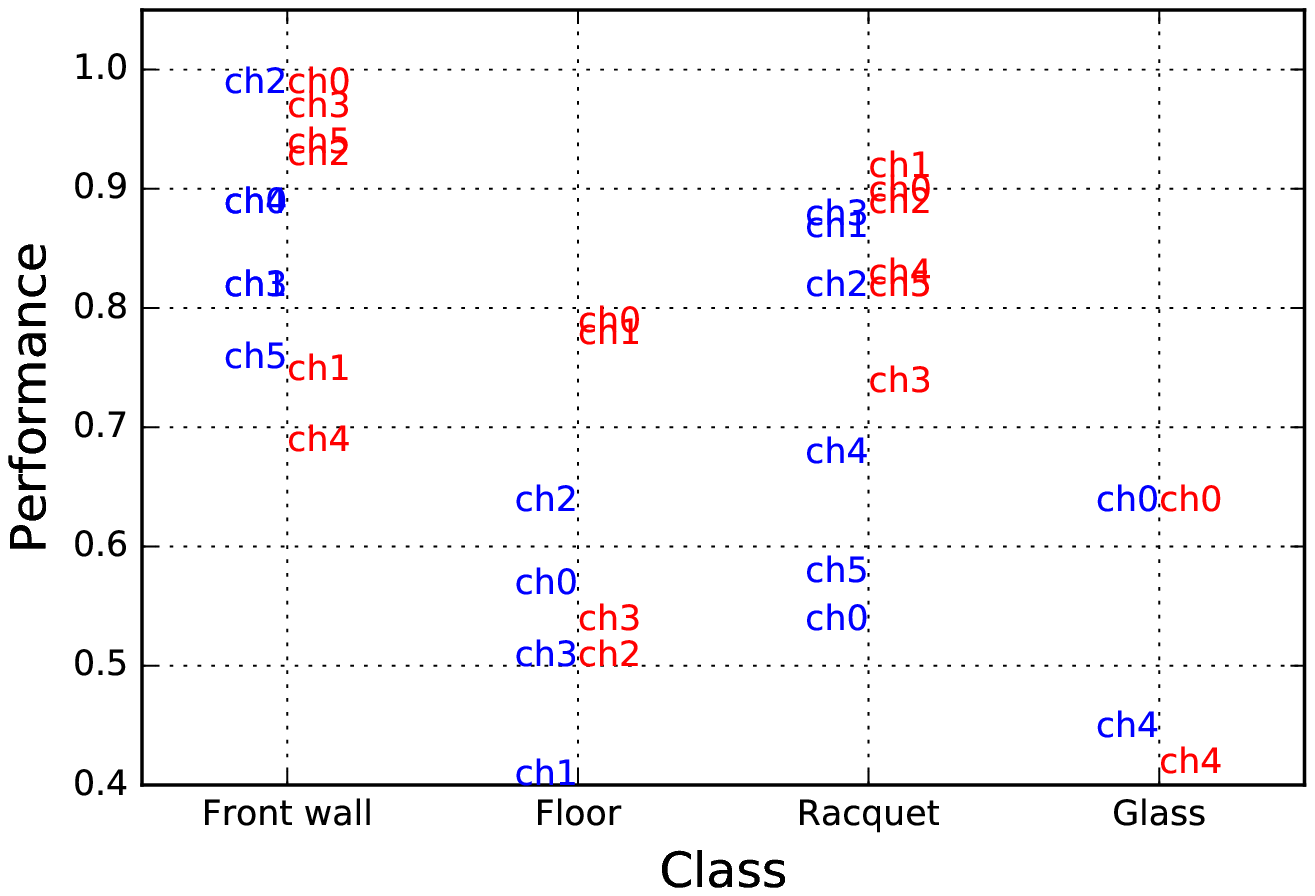}
\caption{{\bf The classifiers' recall.} 
The classwise recall of each channel is presented in $T_1$ (blue) and $T_2$ (red) input sets. 
The performance of front wall classification is reliable. 
The recall of racquet classification is high on channels 1 and 2 in both sets. 
However, the performance of floor and glass classifications is low.}
\label{fig:cl_rec}
\end{figure}

Eight-fold cross-validation~\cite{arlot2010survey} was used on the datasets to evaluate the performance of the classifiers.
Three measures are investigated closer: the accuracy, the precision and the recall.
Accuracy (in Fig~\ref{fig:cl_acc}) is the ratio of correct classifications and the total number of cases examined ($\frac{n_\mathrm{tp} + n_\mathrm{tn}}{n}$).
Precision (in Fig~\ref{fig:cl_prec}) is the fraction constrained to the relevant cases ($\frac{n_\mathrm{tp}}{n_\mathrm{tp} + n_\mathrm{fp}}$).
Recall (in Fig~\ref{fig:cl_rec}) is the fraction of relevant instances that are retrieved ($\frac{n_\mathrm{tp}}{n_\mathrm{tp} + n_\mathrm{fn}}$).

Table~\ref{table:bestresults} summarises the results of the best classifiers for each class. It can be seen that the classification of the front wall and the racquet events is reliable. However, the precision and the recall of floor and glass events are poor. The reason for it is that these classes are under-represented in the data sets.
Whenever $x$, an unseen sample comes, the best classifiers of each class are applied on the new element. The prediction of the class label $\hat{y}$ to which $x$ belongs to is computed by the following formula:
\[
\hat{y}= \left\{
\begin{array}{c l}	
     \underset{k\in C}{\arg\max}\bigg\{\frac{f_k(x) - \mathrm{cut}_k}{1-\mathrm{cut}_k} \frac{\mathrm{prec}_k}{\sum_{i\in C}\mathrm{prec}_i}\bigg\}, & \exists k:f_k(x) > \mathrm{cut}_k \\
     \text{false event}, & \text{otherwise}
\end{array}\right.
\]
where $C$ is the set of class labels without the class of false events and $f_k(x)$, $\mathrm{cut}_k$ and $\mathrm{prec}_k$ are the confidence, the cutoff value and the precision of the best classifier in class $k$ respectively.

\begin{table}[!ht]
\centering
\caption{{\bf The classwise preformance of the best classifiers.}}
\begin{tabular}{|c|c|c|r|r|r|}
\hline
Class & Channel & Input &  Acc & Prec & Rec \\
\hline
Front wall& ch4 & $T_1$ & $0.98$ & $0.93$ & $0.88$ \\
Racquet & ch2 & $T_1$ & $0.94$ & $0.81$ & $0.81$ \\
Floor & ch4 & $T_2$ & $0.88$ & $0.53$ & $0.7$ \\
Glass & ch0 & $T_2$ & $0.88$ & $0.63$ & $0.5$\\
\hline
\end{tabular}
\label{table:bestresults}
\end{table}

\begin{figure}[!h]
\centering
\includegraphics[width=\textwidth]{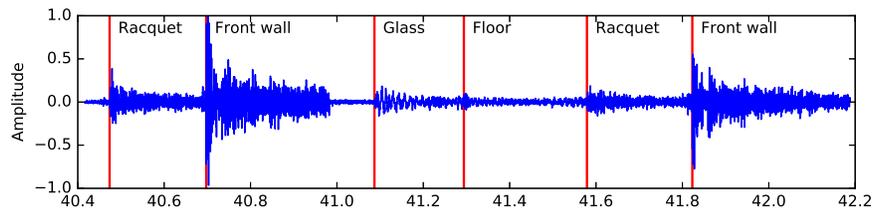}
\caption{{\bf Labelled audio signal.}
1.77~second long samples from channel ch1 in \emph{Audio 2}. 
Detected timestamps and the event classes are marked.}
\label{fig:cl_labelled}
\end{figure}

Fig~\ref{fig:cl_labelled} depicts the combined output generated by the detector and the classifier modules.
A 1.77~seconds long segment of channel 1 audio samples are grabbed from \emph{Audio 2}.
Detections and resolved classes are also shown.
From the snapshot one can observe the different intensities of the events. 
Generally the change in the ball's moment happens when a racquet or a front wall impacts and the sample amplitudes are higher, whereas floor and glass events tend to generate lower intensity and are harder to detect.
 
\subsection{Localization Results}

Based on the geometry of the court, the placement of the microphones and using the localization technique detailed in this study for each set of detection timestamps the 3-d position of the source of the event can be estimated.
In case not all source channels provide a detection of the event localization is still possible. 
Four or more corresponding timestamps will yield a 3-d estimate, whereas with three timestamps the localization of events constrained on a surface (e.g. planes like wall or floor) remains possible.

\begin{figure}[!h]
\centering
\includegraphics[width=.9\textwidth]{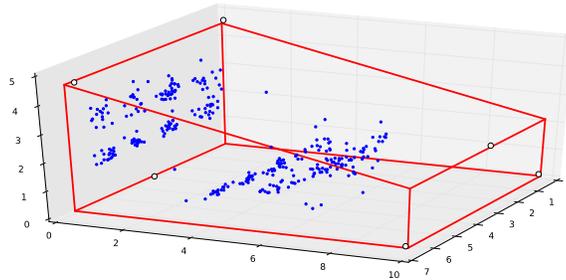}
\caption{{\bf The position of impacts.} Visualize the localized events embedded in 3-d.}
\label{fig:l_3d}
\end{figure}

\begin{figure}[!h]
\centering
\includegraphics[width=0.9\textwidth]{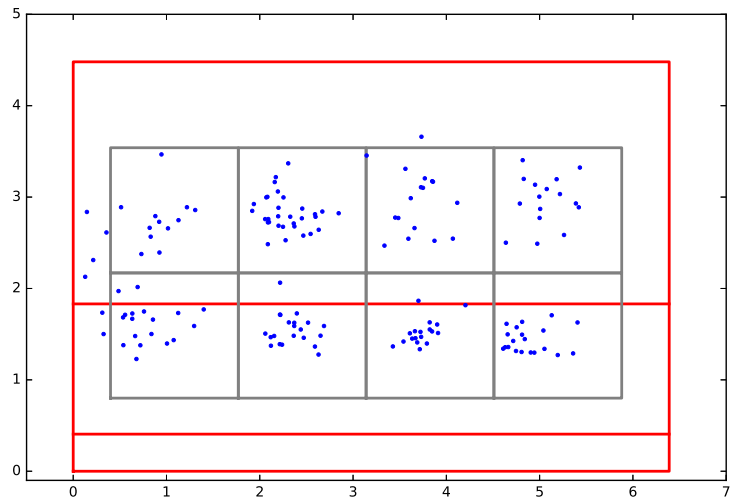}
\caption{{\bf Front wall impacts.} Gray squares embrace the eight target areas.}
\label{fig:l_mwall}
\end{figure}

In Fig~\ref{fig:l_3d} the located events present in dataset \emph{Audio 1} are shown.
In this measurement scenario the player was asked to hit different target areas on the front wall.
It was a rapid exercise, as the ball was shot back at once. 
Only a few times the ball hit the floor, most of the sound is composed of alternating racquet and front wall events.
In Fig~\ref{fig:l_mwall} the front wall events are shown.
The target areas can be seen clearly, and also it is visible the spots scatter a little more on the left.
The reason could be the player being right handed or the fact the target area was hit later during the experiment and the player showed tiredness. 

\begin{figure}[!h]
\centering
\includegraphics[width=0.9\textwidth]{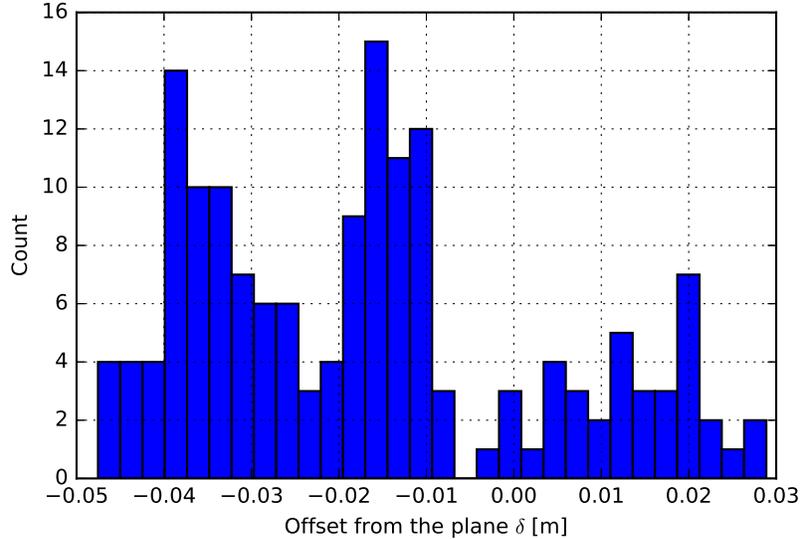}
\caption{{\bf The front wall offsets.} The distribution of the offset $\delta$ from the front wall ($\sigma(\delta)\approx 0.02$~m). }
\label{fig:l_err}
\end{figure}

Measuring the error of the localization method is not straight forward because the ball hitting the main wall does not leave a mark, where the impact happened and there was no means to take pictures of these events.
Taking advantage of the geometry of the front wall an error metric can be defined for front wall events.
The error $\delta$ is defined by the offset of the approximated location from the plane of the front wall.
In Fig~\ref{fig:l_err} the error histogram is shown.
The mean of $\delta$ should vanish and the smaller its variance the better the framework located the events.
From this exercise one can read the standard deviation is $\sigma(\delta)<3$~cm, which is smaller then the size of the squash ball.
 
Another way to define the error is based on relying on human readings of the events.
In the dataset \emph{Audio 1} all of the sound events were marked by human as well as by the detector algorithm.
Localizing the events using both inputs the direct position difference can be investigated.
The mean difference between the positions is 11.8~cm and their standard deviation is 39.9~cm.

\section{Discussion}
\label{Discussion}

Our results support that in sports, where the relevant sound patterns are distinguishable, careful signal processing allows the localisation of shots.
The described system is optimized for handling events and as a consequences the real-time analysis of data is possible, which is important to give an instant feedback. 
The framework can be extended to provide higher level statistics of events such as the evolution of shots types.
From the wide range of possible applications we highlight three use cases.
Firstly, during a match the players can get to know their precision in short time and if is necessary they can change their strategy. 
Secondly, during practice coaches can track the development of the players hit accuracy.
Or thirdly, certain exercises can be defined, which can be automatically and objectively evaluated, without the need for the coach be present during the exercise.

\section{Related work}
\label{sec:relwork}


Squash and soccer were the first sports to be analysed by ways of analysis systems. Formal scientific support for squash emerged at the late 1960s.
The current applications of performance analysis techniques in squash are deeply investigated in the book of Stafford et al.~\cite{obe2016current}.

One test that was developed by squash coach Geoffry Hunt is the ``Hunt Squash Accuracy Test'' (HSAT)~\cite{williams2014measuring}, that is a reliable method used by coaches to assess shot hitting accuracy. 
The test is composed of 375 shots across 13 different types of squash strokes and it is evaluated based on a total score expressed as the number of successful shots.

Recent technological advances have facilitated the development of sport analytical software such as Dartfish video based motion analysis system~\cite{barris2008review,travassos2013performance}. 
However, these systems still require a considerable amount of professional assistance. 

To the best of our knowledge there is no previous research investigating the applicability of sound analysis techniques for squash performance analysis.

\section*{Acknowledgements}

The hardware components enabling this study are installed at Gold Center's squash court. 
We thank them for this opportunity and squash coach Shakeel Khan for the fruitful discussions.

We thank the support of SmartActive project run by Ericsson Hungary Research and Development Center.

\nolinenumbers

\end{document}

%% file: abstract.tex
In competitive sports it is often very hard to quantify the performance.
A player to score or overtake may depend on only millesimal of seconds or millimeters.
In racquet sports like tennis, table tennis and squash many events will occur in a short time duration, whose recording and analysis can help reveal the differences in performance.
In this paper we show that it is possible to architect a framework that utilizes the characteristic sound patterns to precisely classify the types of and localize the positions of these events.
From these basic information the shot types and the ball speed along the trajectories can be estimated.
Comparing these estimates with the optimal speed and target the precision of the shot can be defined.
The detailed shot statistics and precision information significantly enriches and improves data available today.
Feeding them back to the players and the coaches facilitates to describe playing performance objectively and to improve strategy skills.
The framework is implemented, its hardware and software components are installed and tested in a squash court.

%% file: Squash-arxiv.bbl
\begin{thebibliography}{10}

\bibitem{broadbent_tenniscam}
Broadbent DP, Ford PR, O’Hara DA, Williams AM, Causer J.
\newblock The effect of a sequential structure of practice for the training of
  perceptual-cognitive skills in tennis.
\newblock PLOS ONE. 2017;12(3):1--14.
\newblock doi:{10.1371/journal.pone.0174311}.

\bibitem{welfordvariance}
Welford BP.
\newblock Note on a Method for Calculating Corrected Sums of Squares and
  Products.
\newblock Technometrics. 1962;4(3):419--420.

\bibitem{BayesianSurprise}
Boris S, Stiefelhagen R.
\newblock ``Wow!'' Bayesian surprise for salient acoustic event detection.
\newblock 2013 IEEE International Conference on Acoustics, Speech and Signal
  Processing. 2013;.

\bibitem{kullback}
Kullback S, Leibler RA.
\newblock On Information and Sufficiency.
\newblock The Annals of Mathematical Statistics. 1951;22(1):79--86.

\bibitem{Hinton:2012}
Hinton G, et~al.
\newblock Deep neural networks for acoustic modeling in speech recognition: The
  shared views of four research groups.
\newblock IEEE Signal Processing Magazine. 2012;29.6:82--97.

\bibitem{bugatti2002audio}
Bugatti A, Flammini A, Migliorati P.
\newblock Audio classification in speech and music: a comparison between a
  statistical and a neural approach.
\newblock EURASIP Journal on Advances in Signal Processing. 2002;2002(4):1--7.

\bibitem{shao2003applying}
Shao X, Xu C, Kankanhalli MS.
\newblock Applying neural network on the content-based audio classification.
\newblock In: Information, Communications and Signal Processing, 2003 and
  Fourth Pacific Rim Conference on Multimedia. Proceedings of the 2003 Joint
  Conference of the Fourth International Conference on. vol.~3. IEEE; 2003. p.
  1821--1825.

\bibitem{wang2007sound}
Wang Y, Lee CM, Kim DG, Xu Y.
\newblock Sound-quality prediction for nonstationary vehicle interior noise
  based on wavelet pre-processing neural network model.
\newblock Journal of Sound and Vibration. 2007;299(4):933--947.

\bibitem{chawla2002smote}
Chawla NV, Bowyer KW, Hall LO, Kegelmeyer WP.
\newblock SMOTE: synthetic minority over-sampling technique.
\newblock Journal of artificial intelligence research. 2002;16:321--357.

\bibitem{arlot2010survey}
Arlot S, Celisse A, et~al.
\newblock A survey of cross-validation procedures for model selection.
\newblock Statistics surveys. 2010;4:40--79.

\bibitem{obe2016current}
OBE NM.
\newblock Current applications of performance analysis techniques in squash.
\newblock Science of Sport: Squash. 2016;.

\bibitem{williams2014measuring}
Williams BK, Hunt GB, Graham-Smith P, Bourdon PC.
\newblock Measuring squash hitting accuracy using the ‘Hunt squash accuracy
  test’.
\newblock In: ISBS-Conference Proceedings Archive; 2014.

\bibitem{barris2008review}
Barris S, Button C.
\newblock A review of vision-based motion analysis in sport.
\newblock Sports Medicine. 2008;38(12):1025--1043.

\bibitem{travassos2013performance}
Travassos B, Davids K, Ara{\'u}jo D, Esteves PT.
\newblock Performance analysis in team sports: Advances from an Ecological
  Dynamics approach.
\newblock International Journal of Performance Analysis in Sport.
  2013;13(1):83--95.

\end{thebibliography}
